\documentclass[11pt]{article}
 
\usepackage{amsmath,amsthm}
\usepackage{amsthm}
\usepackage{amssymb}
\usepackage{amsfonts}

\usepackage[latin1]{inputenc}

\usepackage{graphicx}

\newtheorem{theorem}{Theorem}

\newcommand{\bq}{\mathbf{q}}
\newcommand{\bp}{\mathbf{p}}
\newcommand{\te}{\theta}

\newcommand{\RR}{\mathbb{R}}
\newcommand{\cM}{{\mathcal M}}
\newcommand\minus\backslash

\newcommand{\la}{{\lambda}}

\newcommand\uno{{\rm I}}
\newcommand\dos{{\rm II}}
\newcommand\tres{{\rm III}}
\newcommand{\rc}{{r_{\rm c}}}

\newcommand\be{\begin{equation}}
\newcommand\ee{\end{equation}}
\newcommand\bea{\begin{eqnarray}}
\newcommand\eea{\end{eqnarray}}

\newcommand{\om}{\omega}
\newcommand\arc{{\rm arc}\!}
\newcommand\dd{{\rm d}}
\newcommand{\bL}{\mathbf{L}}

\begin{document}

 \
 \smallskip

  \noindent
{\LARGE{\bf{On two  superintegrable  nonlinear  oscillators\\[6pt] in N dimensions}}}

\bigskip

\bigskip

\begin{center}
{\large  \'Angel Ballesteros$^a$,   Alberto Enciso$^b$,  Francisco J. Herranz$^a$,\\[4pt] Orlando Ragnisco$^c$ and Danilo Riglioni$^c$}
\end{center}

\noindent
{$^a$ Departamento de F\1sica,  Universidad de Burgos,
09001 Burgos, Spain\\ ~~E-mail: angelb@ubu.es \quad fjherranz@ubu.es\\[10pt]
}
$^b$ Departamento de F\1sica Te\'orica II,   Universidad Complutense,   28040 Madrid,
Spain\\ ~~E-mail: aenciso@fis.ucm.es\\[10pt]
$^c$ Dipartimento di Fisica,   Università di Roma Tre and Istituto Nazionale di
Fisica Nucleare sezione di Roma Tre,  Via Vasca Navale 84,  00146 Roma, Italy  \\
~~E-mail: ragnisco@fis.uniroma3.it  \quad riglioni@fis.uniroma3.it    \\[10pt]

\medskip
\medskip

\begin{abstract}
\noindent 
We consider the classical superintegrable Hamiltonian system given by
$$
{\cal H}_\la={\cal T}+{\cal U}=  \frac{\bp^2}{2(1+\la \bq^2)} +  \frac{ \om^2 \bq^2}{2(1+\la \bq^2)},
$$
where $\cal U$ is known to be the ``intrinsic" oscillator potential on the Darboux spaces of nonconstant curvature determined by the kinetic energy term $\cal T$ and parametrized by $\la$.
We show that ${\cal H}_\la$ is St\"ackel equivalent to the free Euclidean  motion, a fact that directly  provides a curved Fradkin tensor of constants of motion for ${\cal H}_\la$. Furthermore, we analyze in terms of $\la$ the three different underlying manifolds whose geodesic motion is provided by $\cal T$. As a consequence, we find that ${\cal H}_\la$ comprises three different nonlinear physical models that, by constructing their radial effective potentials, are shown to be two different nonlinear oscillators and an infinite barrier potential. The quantization of these two oscillators and its connection with spherical confinement models is briefly discussed.
 \end{abstract}

\bigskip\bigskip\bigskip 


\medskip


\newpage


\section{Introduction}

 Let us consider the $N$-dimensional ($N$D) classical Hamiltonian defined by
  \be
{\cal H}_\la={\cal T}(\bq,\bp)+{\cal U}(\bq)=  \frac{\bp^2}{2(1+\la \bq^2)} +  \frac{ \om^2 \bq^2}{2(1+\la \bq^2)},
 \label{ac}
 \ee
where    $\la$ and $\om$ are real parameters, and   $\bq,\bp\in\RR^N$ are conjugate coordinates and momenta with canonical Poisson  bracket $\{q_i,p_j\}=\delta_{ij}$.

 The mathematical and physical relevance of this system   rely on two main properties~\cite{PhysD}: (i) ${\cal H}_\la$ is a 
maximally superintegrable (MS) Hamiltonian,  since it is endowed with the maximum possible number of  $2N-1$ functionally independent integrals of motion; and (ii)  the central potential ${\cal U}(\bq)$ can be interpreted as the ``intrinsic" oscillator 
on the underlying curved manifold defined through the kinetic term $\cal T$. In particular,   ${\cal T}$   determines the geodesic motion of a particle with unit mass  on a conformally flat space  which was constructed in~\cite{PLB,annals} and is the $N$D spherically symmetric generalization of the Darboux surface of type III~\cite{Ko72,KKMW03}. The corresponding metric  and   scalar curvature depend on $\la$ and are  given by
 \be
 \dd s^2= (1+\la \bq^2)\dd \bq^2 , \qquad  R(\bq)=-\la\,\frac{(N-1)\bigl( 2N+3(N-2)\la \bq^2\bigr)}{(1+\la \bq^2)^3} .
 \label{ad}
 \ee
From this viewpoint, ${\cal H}_\la$ can be regarded  as a  MS   ``$\la$-deformation" of the $N$D   isotropic harmonic oscillator with frequency $\om$ since  $\lim_{\la\to 0} {\cal H}_\la=\frac 12 \bp^2+\frac 12 \om^2\bq^2$.

We recall that ${\cal H}_\la$  can be identified as a particular case within other  frameworks such as: 
(i)  the     ``3D  multifold Kepler"  Hamiltonians~\cite{IK95,uwano} (which   generalize    the MIC--Kepler and   Taub-NUT  systems); (ii)  the ``3D Bertrand systems"~\cite{Perlick,Bertrand,commun} (coming from a generalization of  the classical Bertrand's theorem~\cite{Bertrand2}  to curved spaces); and  (iii) the  ``$N$D
 position-dependent mass  systems"~\cite{Roos,mass2,mass3,mass4,Quesnea,Quesne,Quesnec,ManoloPLA,mass5}  (see also references therein) provided that   the conformal factor of the metric (\ref{ad})  is identified with the variable mass function  
$ m(\bq)=1+\la \bq^2$.
 
The aim of this paper is twofold. On one hand, in the next section we provide a deeper insight in the set of integrals of motion of ${\cal H}_\la$ given in~\cite{PhysD} by applying the so-called St\"ackel transform or coupling constant metamorphosis~\cite{Hietarinta,Stackel2,Stackel4,Sergyeyev,Stackel5}. In this way, we obtain the corresponding $\la$-deformation of the Fradkin tensor of integrals of motion~\cite{Fradkin} for the isotropic harmonic oscillator. On the other hand, we explicitly show that ${\cal H}_\la$ gives rise, in fact, to {\em three  different} physical models. For this latter (and main) purpose, we   present in section 3 which are the underlying manifolds   that come out according to  the values of  $\la$. This analysis leads to {\em three} types of manifolds   which, in turn, correspond to two nonlinear oscillator systems plus a barrier-like one,  which are
studied in section 4  by constructing their associated effective potential.  The  final result is that the Hamiltonian  ${\cal H}_\la$  comprises the {\em hyperbolic}  oscillator   ($\la>0$), the {\em spherical} one (the ``interior" space with $\la<0$) and an infinite potential {\em barrier} (the ``exterior" space with $\la<0$). Remarkably enough, the effective oscillator  potentials are, in this order, hydrogen-like and oscillator-like, which means that the quantization of ${\cal H}_\la$ would provide different types of spherical confinement models like, for instance,~\cite{Jaber,Aquino}. First results in this direction~\cite{ftc} are briefly sketched.


\section{Superintegrability and the St\"ackel transform}

The MS property of   ${\cal H}_\la$ is  characterized by the following statement.
  
\begin{theorem}
(i) The Hamiltonian ${\cal H}_\la$  (\ref{ac}), for any real value of $\la$, is endowed with the following constants of motion.

\noindent
$\bullet$ $(2N-3)$  angular momentum integrals:
\be
  C^{(m)}=\!\! \sum_{1\leq i<j\leq m} \!\!\!\! (q_ip_j-q_jp_i)^2 , \qquad 
 C_{(m)}=\!\!\! \sum_{N-m<i<j\leq N}\!\!\!\!\!\!  (q_ip_j-q_jp_i)^2 ,   \label{af}
 \ee
 where $m=2,\dots,N$ and $C^{(N)}=C_{(N)}$.
 
 \noindent
$\bullet$ $N^2$ integrals  which form the ND curved Fradkin tensor:
 \be
 I_{ij}=p_ip_j-\bigl(2\la  {\cal H}_\la(\bq,\bp)-\om^2\bigr) q_iq_j , 
\label{ag}
\ee
where $ i,j=1,\dots,N$ and 
such that    ${\cal H}_\la=\frac 12 \sum_{i=1}^N I_{ii}$.  

\noindent
(ii) Each of the three  sets $\{{\cal H}_\la,C^{(m)}\}$,  
$\{{\cal H}_\la,C_{(m)}\}$ ($m=2,\dots,N$) and   $\{I_{ii}\}$ ($i=1,\dots,N$) is  formed by $N$ functionally independent functions  in involution.

\noindent
(iii) The set $\{ {\cal H}_\la,C^{(m)}, C_{(m)},  I_{ii} \}$ for $m=2,\dots,N$ with a fixed index $i$    is  constituted  by $2N-1$ functionally independent functions. 
\end{theorem}

A restricted version of this result was proven in~\cite{PhysD}, where only the diagonal integrals $I_{ii}$ and the case $\la>0$ was considered. However, the  same algebraic results do hold for  $\la<0$, and this possibility enable us to get other physical systems different from the one with $\la>0$ that was solved in~\cite{PhysD}. We also remark that the existence of a (curved) Fradkin tensor (\ref{ag}) is what makes ${\cal H}_\la$  (\ref{ac}) a distinguished Hamiltonian, that is, a MS one  which  can be regarded as the ``closest  neighbour of nonconstant curvature" to the harmonic oscillator system, which is obtained in the limit  $\la\to 0$.

It is also worth stressing that  theorem 1 can also be proven by relating  ${\cal H}_\la$  with the  {\em free Euclidean  motion}  through a St\"ackel transform~\cite{Hietarinta,Stackel2,Stackel4,Sergyeyev,Stackel5} as follows.

Let $H$ be an ``initial" Hamiltonian, $H_U$ an ``intermediate" one  and ${\tilde H}$ the ``final" system given by
\be
H=\frac{\bp^2}{\mu(\bq)}+V(\bq),\qquad
H_U=\frac{\bp^2}{\mu(\bq)}+U(\bq),\qquad 
 \tilde H=\frac{H}{U}=\frac{\bp^2}{\tilde\mu(\bq)}+ \tilde V(\bq),
\label{ya}
\ee
such that  
\be
\tilde \mu=\mu U ,\qquad \tilde V= {V}/{U}.
\label{yc}
\ee
Then, each {\em second-order} integral of motion   (symmetry) $S$   of $H$ leads to a new one $\tilde S$ corresponding to $\tilde H$ through an ``intermediate"  symmetry $S_U$ of $H_U$. In particular, if $S$ and $S_U$ are given by 
\be
S=\sum_{i,j=1}^Na^{ij}(\bq)p_ip_j+W(\bq)=S_0+W(\bq)
,\quad S_U=S_0+W_U(\bq),
\label{ye}
\ee
then we get  a second-order symmetry of $\tilde H$ in the form
 \be
 \tilde S=S_0-\frac{W_U}U\, H+\frac 1 U\, H .
 \label{yf}
 \ee
  
 In our case, we consider as the initial Hamiltonian $H$ (\ref{ya}) the free one on the $N$D Euclidean space minus a real constant $\alpha$ (related with $\la$ and $\om$):
 \be
 H=\frac 12 \bp^2-\alpha ,\qquad 2\la\alpha=\om^2 .
 \label{yg}
 \ee
 And our aim is to perform a St\"ackel transform to the  Hamiltonian ${\cal H}_\la$   (\ref{ac}) but written in   ``final" form  as
\be
 {\tilde H}={\cal H}_\la-\alpha=\frac 12\left(  \frac{\bp^2-2\alpha}{1+\la \bq^2}  \right)  .
 \label{yh}
 \ee
 Thus it can be checked that the transformation works provided that 
 \be
 \mu=2,\quad V=-\alpha,\quad \tilde \mu=2(1+\la \bq^2),\quad \tilde V=\frac{-\alpha}{1+\la \bq^2},\quad
 U=(1+\la\bq^2) ,
 \ee
 and the intermediate Hamiltonian is the $N$D istropic harmonic oscillator
 \be
 H_U=\frac 12 \bp^2+\la\bq^2+1.
 \label{yyhh}
 \ee
 
 Next we consider the symmetries $S$  of $H$ (\ref{yg}) which is clearly MS and endowed with $2N-1$ functionally independent functions. Some of them are exactly   (\ref{af}):
 \bea
 && S^{(m)}   =\!\! \sum_{1\leq i<j\leq m} \!\!\!\! (q_ip_j-q_jp_i)^2, \qquad S_0^{(m)}=C^{(m)},\qquad W^{(m)}=0,\nonumber\\[2pt]
  && S_{(m)}   =\!\! \sum_{N-m<i<j\leq N}\!\!\!\!\!\!  (q_ip_j-q_jp_i)^2 ,\qquad S_{0,{(m)}}=C_{(m)},\qquad W_{(m)}=0,\nonumber\\[2pt]
 && S_{ij}=p_ip_j, \qquad S_{0,ij}=p_ip_j ,\qquad W_{ij}=0,
 \eea
where $m=2,\dots,N$ and $i,j=1,\dots N$. The symmetries $S_U$  of  
$H_U$ (\ref{yyhh}) read
\bea
 && S_U^{(m)} \equiv S^{(m)} , \qquad S_0^{(m)}=C^{(m)} , \qquad W_U^{(m)}=0,\nonumber\\[2pt]
  && S_{U,(m)}  \equiv S_{(m)}  ,\qquad S_{0,{(m)}}=C_{(m)},\qquad W_{U,(m)}=0,\nonumber\\[2pt]
 && S_{U,ij}=p_ip_j +2\la q_iq_j,\qquad S_{0,ij}=p_ip_j ,\qquad W_{U,ij}=2\la q_i q_j .
 \eea
Consequently the Hamiltonian $ {\tilde H}$ (\ref{yh}) is also MS and its integrals of motion $\tilde S$  (\ref{yf}) turn out to be
 \bea
 && \tilde S^{(m)}   =\!\! \sum_{1\leq i<j\leq m} \!\!\!\! (q_ip_j-q_jp_i)^2 +  \tilde H=C^{(m)} +\tilde H   ,\nonumber\\[2pt]
  && \tilde S_{(m)}   =\!\! \sum_{N-m<i<j\leq N}\!\!\!\!\!\!  (q_ip_j-q_jp_i)^2 +\tilde H=C_{(m)} +\tilde H  , \nonumber\\[2pt]
 && \tilde S_{ij}=p_ip_j -2\la q_iq_j\tilde H +\tilde H .
 \eea
Finally by introducing $\tilde H={\cal H}_\la-\alpha$ we recover all the results given in theorem 1 proving that, in fact, the Hamiltonian ${\cal H}_\la$ is St\"ackel equivalent to the free Euclidean motion.


 \section{The   underlying Darboux  manifolds}
\label{S.S2}

We recall that the real parameter  $\la=1/\kappa$ was restricted in~\cite{PhysD}  to take a {\em positive} value.  Clearly, the superintegrability properties of the Hamiltonian stated in theorem 1 do hold   for a {\em negative}  $\la$ as well.  Nevertheless  the underlying space and the oscillator potential change dramatically with the sign of $\la$ in such a manner that   
 the domain of the Hamiltonian must be restricted when $\la<0$. 
Hence, the ``generic" Darboux space (that is, the Riemannian manifold with metric (\ref{ad})   determined by the kinetic part of~(\ref{ac})) leads to {\em three} different manifolds $\cM^N$  which have  the following geometric and topological properties.


\subsection{Type  $\uno$: $\la>0$}
The Darboux space is the complete Riemannian manifold $\cM^N=(\RR^N,g)$, with metric 
$g_{ij}:=(1+\la\bq^2)\,\delta_{ij}$. The scalar curvature $R(r)\equiv R(|\bq|)$ (\ref{ad}) is always a  {negative} increasing function such that 
 $\lim_{r\to \infty}R=0$ and  it has a minimum at the origin $$R(0)=-2\la N(N-1),$$ which is exactly the scalar curvature of the $N$D {\em hyperbolic space} with negative constant sectional curvature equal to $-2\la$.
 

\subsection{Type  $\dos$:  $\la<0$ restricted to the {interior space}}

 In this case we consider   the interior   Darboux space defined by $\cM^N=(B_{\rc},g)$ such that 
$$
g_{ij}:=(1-|\la|\bq^2)\,\delta_{ij},\qquad B_\rc=[0,\rc ),\qquad \rc=|\bq|_{\rm c}=1/\sqrt{|\la|},\label{metric2}
$$
that is, $B_\rc$ denotes the ball centered at $0$ of radius $\rc$ which is the critical or singular value for which $R(r)$ diverges and $\lim g_{r\to \rc^-}=0$.
  It is clear that $\cM^N$ is incomplete as a Riemannian manifold.  Notice also that $$R(0)=2|\la| N(N-1),$$ which coincides with the  the scalar curvature of the $N$D {\em spherical space} with positive constant sectional curvature equal to $2|\la|$. 
The behavior of  $R(r)$  depends on the dimension $N$ as follows.

\begin{itemize}

\item  When $2\le N\le 6$,  the scalar curvature  is    a  {positive} increasing function such that  ${\lim_{r\to r^-_{\rm c}} R(r)=+\infty}$.

\item If $7\le N$, there is  a positive maximum for $R(r)$  corresponding to
$$
r_{\rm max}= \sqrt{\frac{N+2}{2(N-2)|\la|}}  , \qquad
R(r_{\rm max})=\frac{ 4|\la|(N-1)(N-2)^3}{(N-6)^2} ,
$$
and ${\lim_{r\to r^-_{\rm c}} R(r)=-\infty}$. 
\end{itemize}


\subsection{Type  $\tres$:   $\la<0$ restricted to the {exterior space}}

To consider the exterior Darboux space,  $\cM^N=(\RR^N\minus\overline{B_{\rc}},g)$, requires to change the sign of both the metric and scalar curvature (\ref{ad}):
\bea
&& g_{ij}:=(|\la|\bq^2-1)\,\delta_{ij},\quad  \RR^N\minus\overline{B_{\rc}}=(\rc ,\infty),\nonumber\\
&&R(\bq)=|\la|\,\frac{(N-1)\bigl( 2N-3(N-2)| \la| \bq^2\bigr)}{(|\la| \bq^2-1)^3} .
\nonumber
\eea
Note that   $\cM^N$ is again incomplete.  According to the dimension $N$, the function $R(r)$ behaves as follows:

\begin{itemize}

\item For $N=2$,   this  is   a  {positive} decreasing function such that  ${\lim_{r\to r^+_{\rm c}} R(r)=+\infty}$ and ${\lim_{r\to \infty} R(r)=0}$.

\item  If $3\le N\le 5$,  the scalar curvature has a negative minimum
$$
r_{\rm min}= \sqrt{\frac{N+2}{2(N-2)|\la|}}  , \qquad
R(r_{\rm min})=-\frac{ 4|\la|(N-1)(N-2)^3}{(N-6)^2} ,
$$
with  ${\lim_{r\to r^+_{\rm c}} R(r)=+\infty}$ and  ${\lim_{r\to \infty} R(r)=0}$.

\item When $6\le N$, $R(r)$ is a negative increasing function with ${\lim_{r\to r^+_{\rm c}} R(r)=-\infty}$ and ${\lim_{r\to \infty} R(r)=0}$.

\end{itemize}


\section{Three radial systems and their effective potentials}

Firstly, we remark that ${\cal H}_\la$ can also be expressed in terms of hyperspherical coordinates $r,\te_j$, 
  and canonical   momenta $p_r,p_{\te_j}$,   $(j=1,\dots,N-1)$  defined by
   \begin{equation}
q_j=r \cos\te_{j}     \prod_{k=1}^{j-1}\sin\te_k ,\quad 1\leq j<N,\qquad 
q_N =r \prod_{k=1}^{N-1}\sin\te_k ,
\label{ba}
\end{equation}
so, $r=|\bq|$. Thus the Hamiltonian      (\ref{ac}) reduces to  a 1D radial system:
   \be
 {\cal H}_\la(r,p_r)= 
 \frac{p_r^2+r^{-2}\bL^2 }{2(1+\la r^2)} +  \frac{ \om^2 r^2}{2(1+\la r^2)} ={\cal T}(r,p_r)+{\cal U}(r) ,
 \label{bg}
 \ee
where  $\bL^2\equiv C^{(N)}\equiv C_{(N)}$ is the total   angular momentum   given by
\begin{equation}
\bL^2=\sum_{j=1}^{N-1}p_{\te_j}^2\prod_{k=1}^{j-1}\frac{1}{\sin^{2}\te_k} .
\label{bf}
\end{equation}
 

Now, the geometric  analysis performed in the previous section indicates that we must deal with {\em three} different physical systems that,  for the types \uno\ and \dos\ we name {\em nonlinear hyperbolic oscillator} and  {\em nonlinear spherical oscillator}, respectively. In these two cases the generic expression for the Hamiltonian (\ref{ac}) is kept (with the boundary $\rc$ for type \dos), while for type \tres\ the sign of the Hamiltonian has to be reversed, thus ensuring a positive kinetic term (and provided that the corresponding restriction on the domain is considered).
In particular, as far as the nonlinear radial potential ${\cal U}(r)$ (\ref{bg})  is concerned we point out   the following facts:

\noindent
$\bullet$    Nonlinear hyperbolic oscillator. When $\la>0$, the potential is   a  {\em positive increasing}  function,   such that 
\be
{\cal U}(r)=  \frac{ \om^2 r^2}{2(1+\la r^2)}     ,\qquad {\cal U}(0)=0 ,\qquad\mbox{and} \qquad  \lim_{r\to \infty}{\cal U}(r)=\frac{\om^2}{2\la} .
\label{pota}
\ee

\noindent
$\bullet$    Nonlinear spherical oscillator. If $\la<0$ and $r<\rc$, the potential is also  a  {\em positive increasing}  function verifying
\be
{\cal U}(r)=  \frac{ \om^2 r^2}{2(1-|\la| r^2)}     ,\qquad {\cal U}(0)=0 ,\qquad\mbox{and} \qquad \lim_{r\to \rc^-}{\cal U}(r)=+\infty .
\label{potb}
\ee

\noindent
$\bullet$   Exterior potential.  When $\la<0$ and $\rc<r$ we impose the change of the sign of the Hamiltonian. In this way the potential becomes  a  {\em positive decreasing}  function: 
 \be
{\cal U}(r)=  \frac{ \om^2 r^2}{2(|\la| r^2-1)}  ,\qquad \lim_{r\to \rc^+}{\cal U}(r)=+\infty    ,\qquad \lim_{r\to \infty}{\cal U}(r)= \frac{\om^2}{2|\la|}.
\label{potc}
\ee
      
But it is essential to stress that each of the above  potentials has to be considered on the corresponding curved space described in section 3. In this respect, the   complete classical system   can be better understood by introducing an effective potential (EP) that takes into account each curved background. This can be achieved by applying a 1D  canonical transformation~\cite{ftc}
$$
P=P(r,p_r),\qquad Q=Q(r),\qquad \{Q,P\}=1,
$$
on the 1D radial Hamiltonian (\ref{bg}) yielding
$$
{\cal H}_\la(Q,P)= \frac 12 P^2+ {\cal U}_{\rm eff}(Q ).
$$
Next we present such an  effective potential for the three abovementioned systems.


\subsection{The nonlinear hyperbolic oscillator}

The 1D  canonical transformation is defined by
\be
P(r,p_r)=\frac{p_r}{\sqrt{1+\la r^2}} ,\qquad Q(r)=\frac 12 r\sqrt{1+\la r^2}+\frac{\arc\sinh(\sqrt{\la}r)}{2\sqrt{\la}},
\label{bbhh}
\ee
which  implies that  $Q(r)$  has a unique (continuously  differentiable) inverse $r(Q)$, on the whole positive semiline, that is, both $r,Q\in [0,\infty)$; note that 
 $\dd Q (r) = \sqrt{1+\la r^2}\dd r$. This transformation  yields  the EP
  \be
  {\cal U}_{\rm eff}(Q(r)  )= \frac{ c_N }{2(1+\la r^2)r^2} + \ \frac{ \om^2 r^2}{2(1+\la r^2)},
 \label{bbgg}
 \ee
 where $c_N\ge 0$ is the value of the integral of motion corresponding to the square of the  total angular momentum $C_{(N)}\equiv \bL^2 $ (\ref{bf}).
Hence the radial motion of the   system can be described as the 1D problem given by the potential $ {\cal U}_{\rm eff}(Q (r))$.

In fact, ${\cal U}_{\rm eff}$  is always positive and it has a minimum located at $r_{\rm min}$ such that
\be
r^2_{\rm min}=\frac{\la c_N+\sqrt{\la^2 c_N^2+\om^2 c_N}}{\om^2},\quad    {\cal U}_{\rm eff}(Q (r_{\min}))=-\la c_N+\sqrt{\la^2 c_N^2+\om^2 c_N} .
\label{yaa}
\ee
Therefore,   $r_{\rm min}$ and $  {\cal U}_{\rm eff}(Q (r_{\min}))$  are, respectively, greater and smaller than those corresponding to the isotropic harmonic oscillator, which are
\be
 \la=0\rightarrow\quad r^2_{\rm min}= { \sqrt{  c_N}}/{\om},\qquad    {\cal U}_{\rm eff}(Q (r_{\min}))= \om \sqrt{  c_N} .
\label{yb}
\ee
This EP has  two
representative limits:
\be
\lim_{r\to 0} {\cal U}_{\rm eff}(Q (r))=+\infty,\qquad \lim_{r\to \infty} {\cal U}_{\rm eff}(Q (r))= {\om^2}/({2\la} ),
\label{bz}
\ee
the  latter being coincident with (\ref{pota}).  Thus, this  EP is hydrogen-like (see fig.~1).

\begin{figure}
   \includegraphics[width=0.70\textwidth]{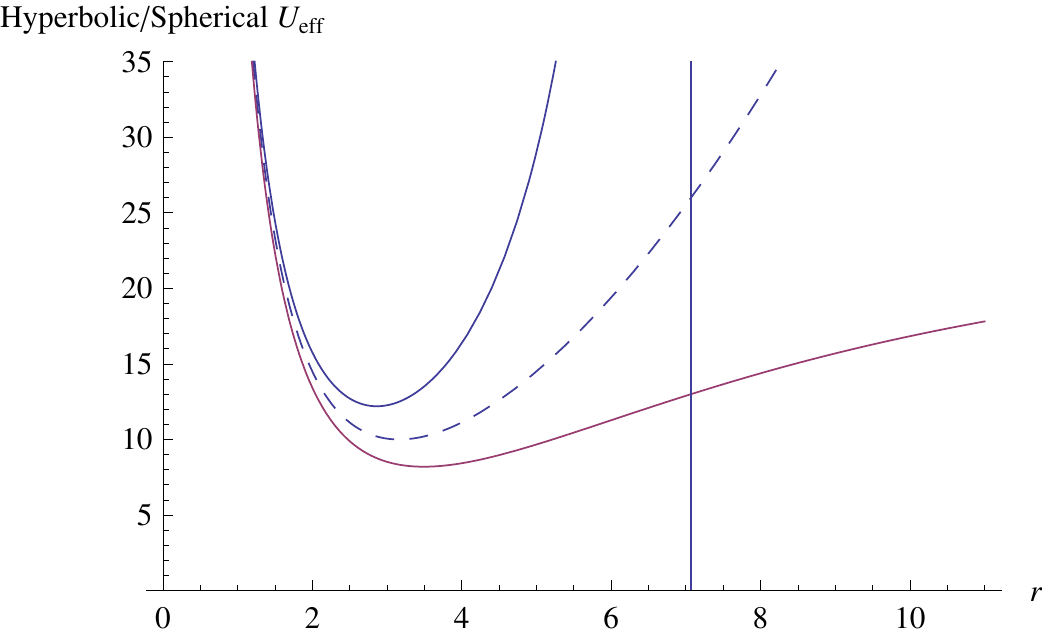}
\caption{The  {effective  nonlinear  hyperbolic and spherical    oscillator potentials}  (\ref{bbgg}) and  (\ref{bbggb})    for    $\la=\pm 0.02$, $c_N=100$ and $\om=1$. The minimum of the hyperbolic potential (red curve) is located at $r_{\rm min}=3.49$ with ${\cal U}_{\rm eff}(r_{\rm min})=8.2$ and ${\cal U}_{\rm eff}(\infty)=25$, while the minimum of the spherical one  (blue curve) is located at $r_{\rm min}=2.86$ with ${\cal U}_{\rm eff}(r_{\rm min})=12.2$ and $\rc=7.07$.   The dashed  line   corresponds to the effective potential of the    harmonic oscillator  with $\la=0$ with minimum ${\cal U}_{\rm eff}(r_{\rm min})=10$ at  $r_{\rm min}=3.16$.}
\label{figU12}        
\end{figure}


\subsection{The nonlinear spherical oscillator}

In this case, the    canonical transformation is given by
\be
P(r,p_r)=\frac{p_r}{\sqrt{1-|\la | r^2}} ,\qquad Q(r)=\frac 12 r\sqrt{1-|\la | r^2}+\frac{\arc\sin(\sqrt{|\la|}r)}{2\sqrt{|\la|}},
\label{bbhhb}
\ee
so that    $Q(r)$  has a unique   inverse $r(Q)$ on the intervals
\be
r\in\left[0,\rc \right),\quad  \rc=\frac{1}{\sqrt{|\la|} };  \qquad Q\in\left[ 0,Q_{\rm c}   \right),\quad Q_{\rm c} =\frac{\pi}{4\sqrt{|\la|} }.
\label{bbz}
\ee
The EP reads
  \be
  {\cal U}_{\rm eff}(Q(r)  )= \frac{ c_N }{2(1-|\la | r^2)r^2} +   \frac{ \om^2 r^2}{2(1-|\la | r^2)},
 \label{bbggb}
 \ee
which  is always positive and it has a minimum located at $r_{\rm min}$ such that
\be
r^2_{\rm min}=\frac{-|\la| c_N+\sqrt{\la^2 c_N^2+\om^2 c_N}}{\om^2},\quad    {\cal U}_{\rm eff}(Q (r_{\min}))=|\la | c_N+\sqrt{\la^2 c_N^2+\om^2 c_N} .
\label{yab}
\ee
But now   $r_{\rm min}$ and $  {\cal U}_{\rm eff}(Q (r_{\min}))$  are, respectively, smaller  and greater  than those corresponding to the isotropic harmonic oscillator (\ref{yb}). This EP has again  two
characteristic limits:
\be
\lim_{r\to 0} {\cal U}_{\rm eff}(Q (r))=+\infty,\qquad \lim_{r\to \rc^-} {\cal U}_{\rm eff}(Q (r))=+\infty ,
\label{bcz}
\ee
which means that we have a deformed oscilator potential that goes smoothly to an infinite barrier as $r$ approaches $\rc^-$ (see fig.~1).


\subsection{The exterior potential}

 The canonical transformation  for the third system turns out to be
 \be
P(r,p_r)=\frac{p_r}{\sqrt{|\la | r^2-1}} ,\quad Q(r)=\frac 12 r\sqrt{|\la | r^2-1}-\frac{\ln\left (2\left( {|\la|}r+\sqrt{|\la|}\sqrt{|\la | r^2-1}\right)  \right )}{2\sqrt{|\la|}}
\label{bbhhc}
\ee
and $Q(r)$  has a unique   inverse $r(Q)$ on the intervals
\be
r\in\left[\rc,\infty\right),\quad \rc=\frac{1}{\sqrt{|\la|} };\qquad Q\in\left[ Q_{\rm c}   ,\infty \right),\quad Q_{\rm c} =- \frac{\ln\left (2 ( \sqrt{|\la|}r\right) }{2\sqrt{|\la|} } .
\label{bbzc}
\ee
The EP is
  \be
  {\cal U}_{\rm eff}(Q(r)  )= \frac{ c_N }{2(|\la | r^2-1)r^2} +   \frac{ \om^2 r^2}{2(|\la | r^2-1)}.
 \label{bbggc}
 \ee
  The function  $ {\cal U}_{\rm eff}$  is again positive but,  unlike   the two previous systems, it has no  minimum; this fulfils the same limits (\ref{potc}) so EP is an infinite (left) potential barrier  which is represented in fig.~2.

\begin{figure}
  \includegraphics[width=0.7\textwidth]{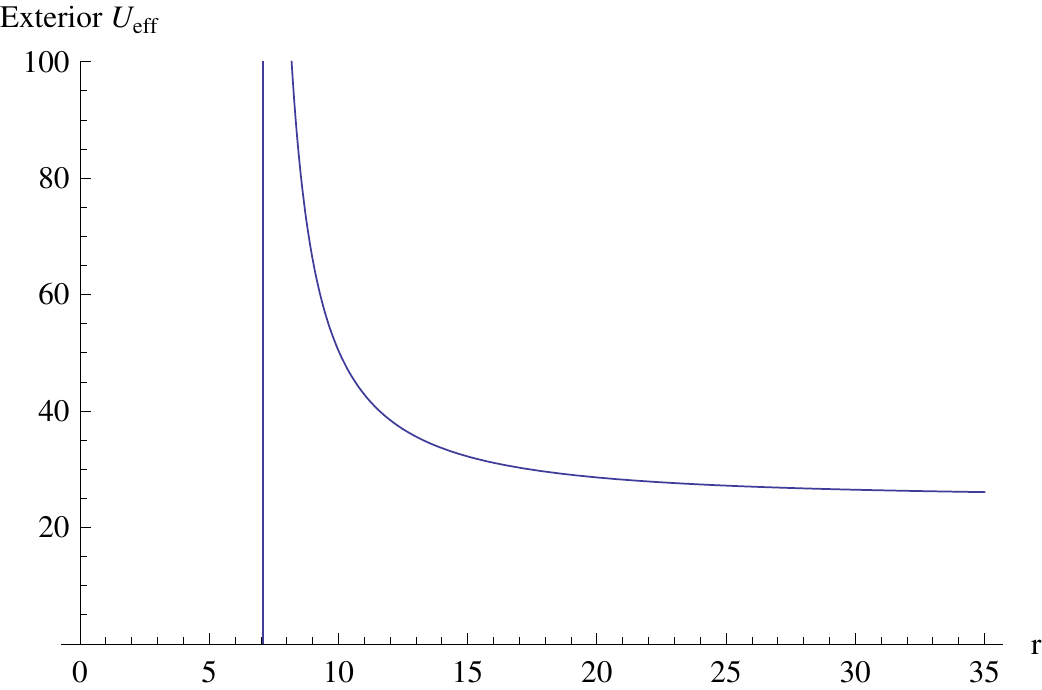}
\caption{The  {effective  nonlinear exterior oscillator potential} (\ref{bbggc}) for    $\la=-0.02$, $c_N=100$ and $\om=1$. The critical point is 
$\rc=7.07$ and  ${\cal U}_{\rm eff}(\infty)=25$}
\label{figU3}       
\end{figure}


Finally, some remarks concerning the quantization of these systems are in order.
The nonlinear hyperbolic oscillator $(\la >0)$ has been fully quantized in~\cite{ftc} and its discrete 
spectrum is given by
\begin{equation}
E_n= -\hbar^2 \lambda\left(n + \frac{N}{2}\right)^2 + \hbar \left(n+\frac N 2\right ) \sqrt{\hbar^2 \lambda^2 \left(n+\frac{N}{2}\right)^2+ \om^2 } .
\label{spectrum}
\end{equation}
The corresponding stationary states have been obtained in analytic form. Note that the limit $n\to\infty$ of $E_n$ is just the asymptotic value ${\om^2}/{2\la} $, as expected.

In view of the shape of the effective potential (see~fig. 1), the quantum spherical oscillator  $(\la <0)$ should provide a new radial confinement model that could be useful as a position-dependent-mass model for spherical quantum dots~\cite{Gritsev}. The exact solution of the corresponding Schr\"odinger problem is still in progress.


\section*{Acknowledgments}

This work was partially supported by the Spanish MICINN   under grants    MTM2010-18556   and FIS2008-00209, by the    Junta de Castilla y
Le\'on  (project GR224), by the Banco Santander--UCM 
(grant GR58/08-910556)
   and by  the Italian--Spanish INFN--MICINN (project ACI2009-1083). F.J.H. is very grateful to W. Miller for helpful suggestions on the St\"ackel transform.

\end{document}